\documentclass[finalversion,simpleeqnnos]{yrl}
\usepackage{times}
\usepackage{graphicx,psfig,epsfig}

\ifx\pdfannotlink\dummy   
\usepackage[hypertex]{hyperref} 
\else
\usepackage[pdftex]{hyperref}   
\fi
\title{Bid Optimization for Internet Graphical Ad Auction Systems via
Special Ordered Sets.
}  
\author{Ralphe Wiggins and John A. Tomlin\\
  2821 Mission College Blvd.\\
  Santa Clara, CA 95054\\
        \{ralphe@wiggins.name, tomlin@yahoo-inc.com\} }
\date{April 30, 2007}      
\yrlnumber{YR-2007-004}

\begin{document}             

\maketitle                   

\begin{abstract} 
This paper describes an optimization model for setting bid levels for certain types of advertisements on web pages. This model is non-convex, but we are able to obtain optimal or near-optimal solutions rapidly using branch and cut open-source software. The financial benefits obtained using the prototype system have been substantial.
\end{abstract}

\section{Introduction}

Advertising on the World Wide Web is ubiquitous and a big business. Recently a great deal of attention has been paid to ``Sponsored Search'', where text advertisements with hypertext links are placed next to the search results produced by search engines (see \cite{Sch}), and these do indeed account for a large fraction of the revenue generated by web advertising. However, a comparable amount of revenue is currently generated by the more traditional graphical advertisements (or banner ads, or more simply, ads) placed on web pages, and that is the type considered in this paper. 

We will frequently refer to an {\em Ad Server}. This is a machine, or set of machines, which receives HTTP requests for ads when a page is viewed, makes a decision on which ad or ads to display in the ad {\em positions}, and returns the ads to the users browser. Since the pages are known as {\em properties}, an ad is said to be allocated to a {\em property/position}. The ad is also said to have received an {\em impression}. The algorithms and data which the ad server uses to decide on which ads are to be shown, in which property/position, are clearly critical to revenue and profitability.

There are two types of graphical ads, or more precisely, {\em ad campaigns} commonly offered by internet companies. The first of these we may call {\em guaranteed} ad campaigns, where the ads are sold for a negotiated price to advertisers, and the ad space (inventory) purchased is guaranteed to be available. The second type are those sold by {\em auction}, and shown on a ``best effort'' basis.  The auctioned ads may be further divided into {\em House Ads}, which an internet company purchases for its own use, and the remainder, which are bought by outside advertisers. It is the House ads that will be our major focus here. 

One of the problems facing companies which act as publisher, content provider and advertising medium is how to divide the inventory that is up for auction between {\em House businesses} and paying clients.  House businesses may contribute to gross income in 3 ways:
\begin{enumerate}
\item	Businesses that sell a product or service that contribute directly to income.  The competition with paying clients for inventory resources should be based on net income minus life time revenue reduced by an estimate of the cost of the business.  
\item	Businesses that enhance traffic by encouraging people to go to other parts of the network.  To the extent that the cost of the clicked-on ads is less than the income on the target property, the net income is the difference between these two amounts and such ads can compete on that basis.  
\item	Finally, there are those businesses that enhance income simply by increasing the appeal of visiting the company's network.  While such "appeal" cannot be accurately quantified, we can use total traffic on each property and a value per user as surrogates.
\end{enumerate}
So long as House businesses can profitably compete for ads under revenue types 1 or 2, then in principle, and all other things being equal, they should have unlimited budgets, since the more they spend, the more the company makes. However, in reality budgets are not unlimited, and we must accept them as constraints. In addition there are other factors, such as ad fatigue\footnote{The phenomenon whereby ads may become less effective the more they are shown to a user}, and the need to deliver the guaranteed ads, which limit such spending.

Note that businesses, including House businesses, may have multiple campaigns which share the same badget. Each campaign is associated with a property/position.

We consider setting, or rather re-setting, bids for house ads in such a way as to (approximately) maximize expected return for a large group of ads. This involves observing the bid levels of other groups of ads and then re-computing our bids in the relevant auctions in such a way as to maximize expected return for the model horizon. This requires choosing among discrete bids, which can be modeled as Special Ordered Sets\cite{SOS} of type 1. A combination of heuristics applied to the LP solution, cutting planes, and branching leads to rapid solution of this discrete model.
 
Use of optimization models for ad campaign planning in the presence of budgets is not completely new, either in the traditional media (see \cite{NBC}) or in the web context. In the sponsored search setting, Abrams et al.\cite{AMT} use a linear programming model with column generation to approach the problem. Several papers have attacked the problem of optimizing ad serving in what we have called the guaranteed environment (see \cite{Abe} and it's references). However, the model discussed here has novel features that require a different approach. The 
models discussed in \cite{Abe} and elsewhere assume that the model output can specify the actual serving of specific ads for a page view---that is dictate a serving policy to the ad server. However, in the non-gauranteed setting, our only ``handle'' may be the bid levels to be set for the auction procedure which in turn affect the serving policy of the ad server. Clearly this implies some form of discrete model, since a bid either exceeds another bid, or it does not, and the outcome will depend on these relative bid levels.

In the remainder of this paper we will outline some of the relevant aspects of ad server behavior, discuss the data available, formulate our model, propose some solution strategies and give computational experience.

\section{Ad Server Behavior}

Ad servers may have quite complicated behavior. For the purposes of this paper it is sufficient to point out that when a request for ads from a page view arrive, a number of factors may be considered. Firstly, the candidate ads may be filtered by a user profile requested by the advertiser, which must be compared with the profile of the user (if any). Secondly, guaranteed ads will usually be shown in preference to non-guaranteed, provided that other factors such as ad fatigue do not interfere. Finally the display of non-guaranteed ads may be determined not only by who ``wins'' the auction for this property/position, but budget limit and profile matching. Thus the ad corresponding to the highest bidder may not always the one shown. When an ad (and in particular a non-guaranteed ad) is shown, the appropriate budget is decremented.

We see from this abbreviated list of characteristics that determining the number of impressions that an ad will receive is not a deterministic  function of the bids alone. We are therefore reduced to approximating the ``value'' of an ad with respect to its bid level by using historical data.
%
%
\begin{figure}[htfb]
\begin{center}
\hspace{0.0cm}
\psfig{figure=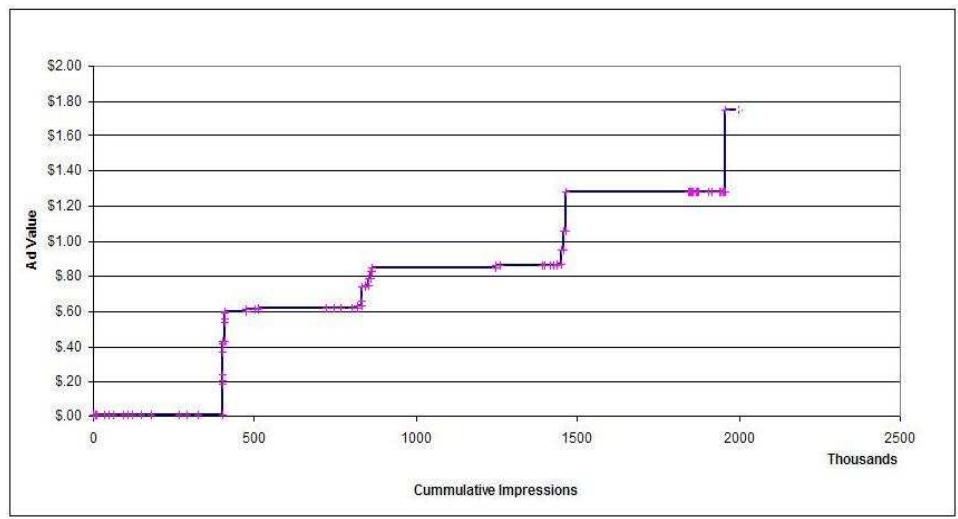,height=2.7in,width=5in}
\caption{Ad Value v. Impressions, Example 1.} \label{fig-1}
\end{center}
\end{figure}
\section{Ad Value, Return, and Impressions}\label{ad-val}

For each ad campaign $i$ we define a discrete set of {\em bid levels} $b_{ij}$, designed to just exceed the (known) competing bids.
A key concept in our model is the {\em ad value} $A_{ij}$ associated with each bid level $j$ for campaign $i$. This is our proxy for the expected amount by which the appropriate budget will be decremented when it gets an impression (i.e. is served up by the ad server). Associated with this ad value is an expected total gross return $L_{ij}$ for making a bid at level $j$ for campaign $i$.

The number of impressions obtained will clearly be influenced by the bid level $j$.  Our model requires that we have a relationship between the ad value and return for the campaign and the number of impressions expected corresponding to the ad values. A real example of such a relationship is shown in Figure \ref{fig-1}. The x-axis value at the right of each horizontal segment is the expected number of impressions received for the bid level associated with the ad value on the y-axis\footnote{The small hash marks represent actual historical bids and volumes}. These values, denoted $P_{ij}$, along with the $A_{ij}$ and $L_{ij}$ are extrapolated from historical data. Note that the actual bids $b_{ij}$ do not appear themselves in the model below, and we do not discuss the derivation of the ad values, returns, and impression counts further in this paper.

\begin{figure}[htfb]
\begin{center}
\hspace{0.0cm}
\psfig{figure=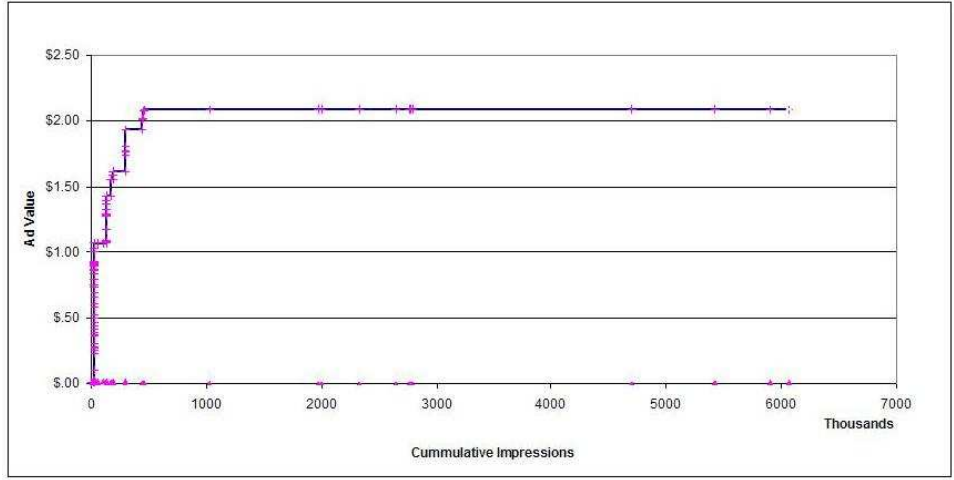,width=5in}
\caption{Ad Value v. Impressions, Example 2.} \label{fig-2}
\end{center}
\end{figure}

\begin{figure}[htfb]
\begin{center}
\hspace{0.0cm}
\psfig{figure=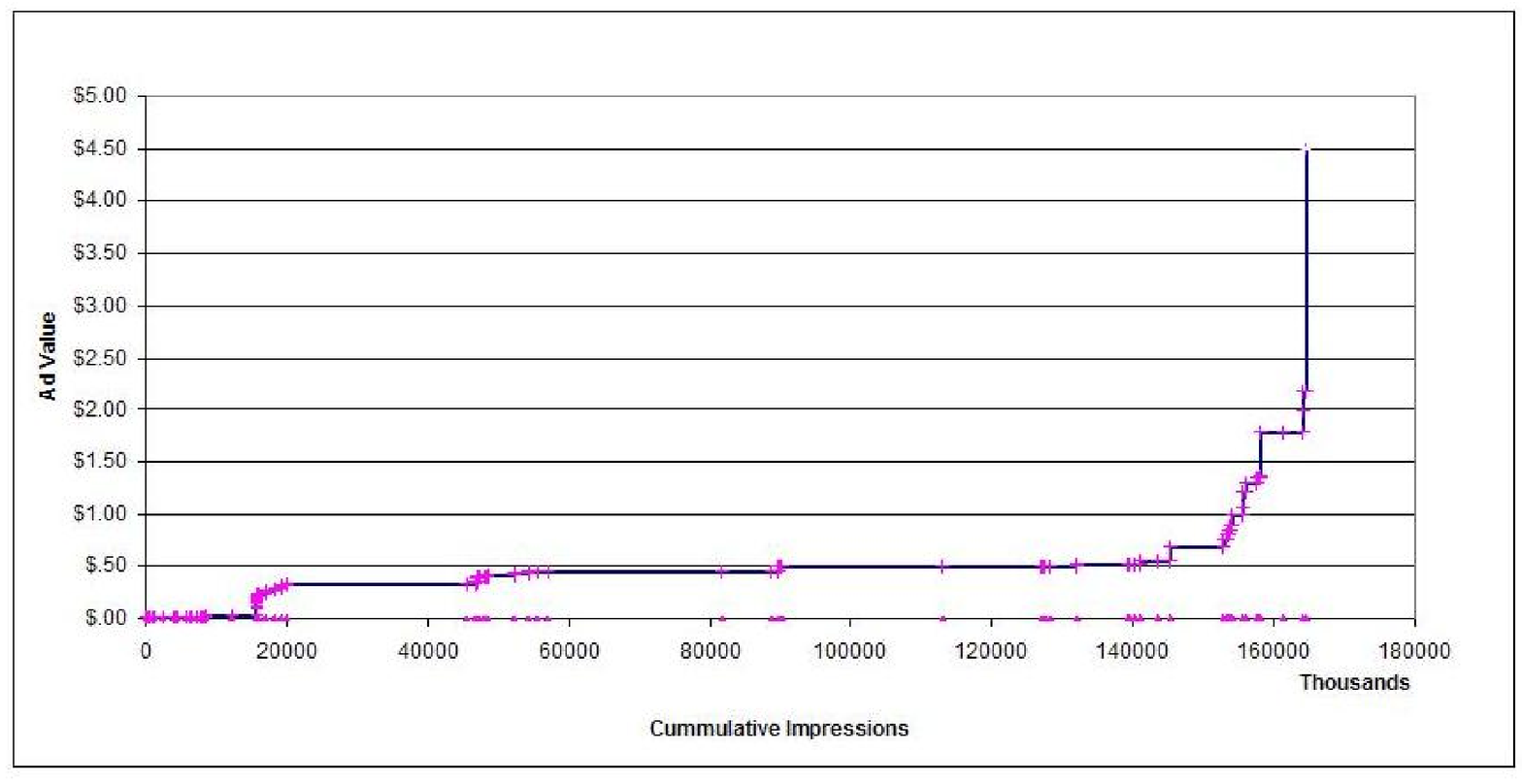,width=5in}
\caption{Ad Value v. Impressions, Example 3.} \label{fig-3}
\end{center}
\end{figure}

The ad value versus impression graphs are not always as regular as that displayed in Figure \ref{fig-1}. More ``lopsided'' examples are shown in Figures \ref{fig-2} and \ref{fig-3}.
%
%

\section{Model Formulation}   

We now formally define the optimization model to be solved:

\vskip 5pt
{\sl Indices} \nopagebreak
\vskip 5pt
   \begin{tabular}{ll}
     $i = 1,...,I$ & The house ad campaigns.
   \\$j = 1,...,J_i$ &  The bid estimate levels for campaign $i$
   \\$k = 1,...,K$ &  The businesses
   \end{tabular}
\vskip 5pt
{\sl Data}  \nopagebreak
\vskip 5pt
   \begin{tabular}{ll}
    $ I_{k} = \{i_1, \ldots, i_{m_k}\}$    & The set of campaigns for business $k$.
     \\ $B_k$ & Budget for business $k$
   \\$L_{ij}$ & Return for campaign $i$, estimate $j$
   \\ $AV_{ij}$ & Ad value associated with pair $(i,j)$
   \\ $P_{ij}$ & Number of impressions for combination $(i,j)$
   \\ $CTR_{ik}$ & Expected click-through rate for business $k$ on campaign $i$
   \\ $CPC_k$ & (Given) cost per click for business $k$
   \\$V$ & Overall impression budget for House ads
\end{tabular}
\vskip 5pt
{\sl Variables}
\vskip 5pt
   \begin{tabular}{ll}
      $\delta_{ij}$   &  has value 1 if level $j$ chosen for campaign $i$, 0 otherwise
   \end{tabular}
\vskip 5pt

{\sl Constraints}
\vskip 5pt
SOS1 (multiple choice)
 \begin{equation}\label{conv}
	\sum_{j} \delta_{ij} = 1 \ \ \ \ \ \ \forall i
 \end{equation}
\vskip 5pt
Budgets
 \begin{equation}\label{budget}
     \sum_{i\in I_k}\sum_j P_{ij}\cdot AV_{ij} \delta_{ij} \le B_{k} \ \ \ \ 
     \forall k
 \end{equation}
\vskip 5pt
Click Values
 \begin{equation}\label{clicks}
     \sum_{i\in I_k}\sum_j P_{ij}\cdot AV_{ij} \delta_{ij} \le 
     CPC_k \cdot \sum_{i\in I_k}\sum_j CTR_{ik} \cdot P_{ij} \delta_{ij}\ \ \
     \forall k
 \end{equation}
\vskip 5pt
Impression Budget
 \begin{equation}\label{i-budget}
     \sum_k \sum_{i\in I_k}\sum_j P_{ij}\delta_{ij} \le V
 \end{equation}

\vskip 5pt

Objective
\vskip 5pt
   \begin{tabular}{ll}
    {\rm Maximize} &$ \ \sum_k\sum_{i\in I_k}\sum_j L_{ij} \delta_{ij}$
   \end{tabular}
\vskip 10pt

The purpose of the $\delta_{ij}$ variables is to choose a bid level from the finite set of possibilities presented for campaign $i$. Following the description in section \ref{ad-val}, we see that the ad value for an insert line $i$ is therefore $\sum_j AV_{ij}\delta_{ij}$, the return is $\sum_j L_{ij}\delta_{ij}$ and the expected number of impressions is $\sum_j P_{ij}\delta_{ij}$. We also insert a ``do nothing'' variable $\delta_{i0}$, or explicit slack, at the beginning of each set. Note that by definition at most one of variables which make up a SOS1 (special ordered set \cite{SOS} of type 1) may be nonzero, and in this case the constraint (\ref{conv}) implies that the SOS1 variables must be zero or one in a valid solution, and therefore integer.

Note that except for the overall impression constraint (\ref{i-budget}), the model falls into disjoint sub models, one for each business $k$. This loose connection makes the model somewhat easier to solve than we might expect for a non-convex model, especially since it may often be non-binding. This would allow solution of a sequence of independent models. However, the later case is hard to predict {\em a priori} and in any case the size of the overall model has so far proved quite manageable.

\section{Implementation and Solution Strategy}

The model we have implemented is generated and solved using a suite of programs. The data on the advertising campaigns and budgets are retrieved from a commercial data base via an SQL program, which feeds them to a C program that generates a standard MPS data file. This is read by the solver, which is built on the COIN-OR open-source C++ library\cite{coin1}. In particular we use the Special Ordered Set capabilities of the Coin Branch-and-Cut (CBC) library\cite{coin2}, using a strategy to be discussed below. When a satisfactory solution is obtained it is written to file in pseudo-MPS output format, for use by another C program which interprets the solution for the bidding software.

One advantage of using this implementation strategy is that it is very easy to design solution strategies which limit the branch and cut search. Since we have introduced several layers of approximation in the formulation of the model, and the derivation of its data, it would be foolish to insist on achieving an exact optimum. Thus a first feasible integer solution is perfectly adequate, provided the integer ``gap'' is small enough, and we may hope to use simple heuristics to give the search a hot start. 

Some familiarity with branch and bound, branch and cut and special ordered sets will be assumed in the remainder of this section, but the reader who is only interested in the results can skip to section \ref{pract}.

We experimented with 2 hot start strategies:
\begin{enumerate}
\item When an SOS is exactly, or almost, satisfied in the LP solution , i.e. one member of the set is close to 1, which is most of the time, that member is fixed to 1, and the other members fixed to zero, provided either that (a) this member is the first member of the set, or (b) the reduced costs of the other members are greater than the some tolerance.
\item If exactly one member of a set is nonzero, it is fixed to 1 and all other members to zero. Otherwise all members up to the first nonzero, and after the last nonzero are fixed to zero.
\end{enumerate} 
There is a slight possibility that these variable fixing strategies will make the problem integer infeasible, in which case we would have to relax them again. We return to this point later on.

In addition to this fixing of variables we apply 3 of the types of ``cuts'' available in the CBC library---known as ``Probing'', ``Gomory'' and ``Knapsack'' cuts. If strategy 2 is used we also add ``Redsplit'' and ``Clique'' cuts.

\begin{table}[htbp]
\begin{centering}
\begin{tabular}{|| c | c | c | c | c ||}
\hline
 & Number & Strategy1 & Time & Best known \\ \hline
Model & of SOS &\% degradation & (seconds) & \% degradation \\ \hline
1 & 2704 & 2.04   & 0.239& 2.04 \\ \hline
2 & 5508 & 2.94 & 10.33 & 2.87 \\ \hline 
3 & 6589 & 6.91 & 0.572 & 6.91 \\ \hline 
4 & 8410 & 18.94 & 0.721 & 18.94 \\ \hline 
5 & 11504 & 6.99 & 44.99 & 6.99 \\ \hline 
6 & 16259 & ???? & $>$1200 & ???? \\ \hline 
\end{tabular}
\caption{Degradations of first integer solutions, Strategy 1}
\label{strat1}
\end{centering}
\end{table}

Tables \ref{strat1} and \ref{strat2} give the results of running some representative problems with the two strategies. The results are given in terms of percentage degradation of the first integer solution found from the continuous LP solution, and the times taken on an Intel Linux box (with Xeon 2.8 GHz processor and   2 GB of RAM). The ``best'' known solution is that found within 1200 seconds. In most cases, we were able to prove optimality of the first solution, subject to the variable fixing that had been carried out (hence the slight differences in the best known solutions for the 2 strategies). However, Strategy 1 was clearly not satisfactory for problem 6, though it solved easily with Strategy 2.

\begin{table}[htbp]
\begin{centering}
\begin{tabular}{|| c | c | c | c | c ||}
\hline
 & Number & Strategy2 & Time & Best known \\ \hline
Model & of SOS &\% degradation & (seconds) & \% degradation \\ \hline
1 & 2704 & 2.167 & 0.107 & 2.167 \\ \hline
2 & 5508 & 3.035 & 0.215 & 3.035 \\ \hline 
3 & 6589 & 6.929 & 0.268 & 6.929 \\ \hline 
4 & 8410 & 19.345 & 0.356 & 19.345 \\ \hline 
5 & 11504 & 9.062 & 0.455 & 9.062 \\ \hline 
6 & 16259 & 8.346 & 8.214 & 8.364 \\ \hline 
\end{tabular}
\caption{Degradations of first integer solutions, Strategy 2}
\label{strat2}
\end{centering}
\end{table}

In general we conclude that both strategies may become too aggressive as models become larger and more complex. Even if the times are acceptable (we expect to solve this daily, or at most hourly), the degradations can become poor.

\section{Relaxation to SOS2}

One approach to the problems seen above is to relax the model. If we consider the relationships expressed in Figures \ref{fig-1}--\ref{fig-3} we see that there is no advantage to having an ad value on the vertical segments of the graphs, unless this allows us to maintain budget feasibility. Maintaining this feasibility is the biggest cause of degradation from the LP solution, furthermore experience shows that this is an issue in only a tiny fraction of the lines in the model. We therefore Cavalierly dispense with the SOS1 requirement that only one member of a set be non-zero, but allow at most two members of the set to be nonzero, and then only if they are adjacent---in other words relax the SOS1 set to a SOS2 set (see \cite{F-T}). This will have no effect for most of the sets, but allow us to ``fudge'' borderline cases.

Following this relaxation, we adopt a simpler hot start procedure for the now non-convex (not integer) tree search. Firstly, the integer cuts must be dispensed with. Secondly, our variable fixing procedure simply looks at the LP solution, and for each set, flags to zero those variables before the first non-zero member, and those after the last non-zero member. If a set is not satisfied we attempt a temporary fixing of all the variables not so far fixed, except the two which define the ``current interval'' as defined in \cite{SOS},\cite{F-T}. The LP is then resolved. If it is feasible this process will have led to a valid---one hopes, good---solution, which may be used to put a bound on the valid solutions. If not, we obtain no such bound. The variables which were temporarily fixed are now unfixed, and we proceed to the branch and bound algorithm.
This strategy, which we call Strategy 3, has been more consistent than use of SOS1, and the one we use in practice. Results for the same set of problems as in Table \ref{strat1} are shown in Table \ref{strat3}. In the rare cases when an SOS2 set is satisfied with 2 non-zero members we simply use the interpolated bid.

Because of the relaxation, the degradations are significantly smaller than with SOS1, as expected, but this should not be considered very significant.

\begin{table}[htbp]
\begin{centering}
\begin{tabular}{|| c | c | c | c | c ||}
\hline
 & Number & Strategy3 & Time & Best known \\ \hline
Model & of SOS &\% degradation & (seconds) & \% degradation \\ \hline
1 & 2704 & 0.270 & 0.809 & 0.270 \\ \hline
2 & 5508 & 0.561 & 1.203 & 0.561 \\ \hline 
3 & 6589 & 0.126 & 1.834 & 0.126 \\ \hline 
4 & 8410 & 0.038 & 3.218 & 0.038 \\ \hline 
5 & 11504 & 0 & 3.958 & 0 \\ \hline 
6 & 16259 & 0.500 & 113.2 & 0.498 \\ \hline 
\end{tabular}
\caption{Degradations of first SOS2 solutions, Strategy 3}
\label{strat3}
\end{centering}
\end{table}

\section{Practical Results}\label{pract}

In our company, use of the model, as opposed to continuing with the traditional manual process, is on an opt-in basis for each House business. It is gratifying that more and more of these businesses have opted in, but perhaps not surprising, since the model attempts to optimize the portfolio of campaigns for each business, rather than treating them independently and greedily.
Without giving company confidential dollar figures, we can indicate the growing practical success of our model by comparing the Return On Investment (ROI) achieved by the businesses which have opted in compared with those that have not. ROI is computed as the ratio of the imputed income to the delivery cost, minus 1. The ROIs achieved by the model and the manual process are shown in Table \ref{adopt} for the period Q4 2005 through Q1 2007.

\begin{table}[htbp]
\begin{centering}
\begin{tabular}{|| l | c | c ||}
\hline
 & Model  & Manual \\ \hline
Quarter & ROI & ROI \\ \hline
Q4 2005 & 0.90  &  0.59 \\ \hline
Q1 2006 & 1.36 & 0.89 \\ \hline 
Q2 2006 & 1.98 & 0.37 \\ \hline 
Q3 2006 & 0.59 & 0.34 \\ \hline 
Q4 2006 & 1.72 & -0.28 \\ \hline 
Q1 2006 & 1.39 & -0.39 \\ \hline
Average & 1.32 & 0.25 \\ \hline 
\end{tabular}
\caption{ROI for Model v. Manual}
\label{adopt}
\end{centering}
\end{table}

The imputed corporate gross income derived from using the model, over this period, is well over 8 figures.

\section{Extensions and Future Work}

The model we have described is clearly capable of being used more widely, and of being extended. An obvious possibility is use of the model by non-House businesses to optimize their portfolio of non-guaranteed ads. This raises the interesting research question of what happens when many competing advertisers use the same model and historical data, a question we leave for the future.

Mathematical extensions of the model are also clearly possible. One such extension would be to specifying particular property/positions in the model, as opposed to using a single ad value. Another would be to specify more complex budgetary and/or impression constraints, perhaps at several levels.

\section{Conclusion}

The problem of setting bid levels for ad campaigns can not only be expressed in terms of a non-convex optimization problem, but efficiently solved to satisfactory accuracy, and the solutions implemented by an ad server which accepts such bids as a basis for its serving decisions. This enables us to increase both efficiency of the process and profitability of the outcome.

\section{Acknowledgements}
We are indebted to our colleagues Ryan Christensen, Andrea Ford, and Madhu Vudali for their encouragement and cooperation in the course of this work.

\bibliographystyle{abbrv}
\bibliography{gekkox}

\end{document}